\documentclass[usenatbib]{mn2e}
\usepackage{lineno}
\usepackage{natbib}
\usepackage{psfig}
\usepackage{graphicx}
\usepackage{amssymb}
\usepackage{color}

\author[Perera et al.]{B.~B.~P.~Perera$^{1}$, B.~W.~Stappers$^{1}$, P.~Weltevrede$^{1}$, A.~G.~Lyne$^{1}$, C.~G.~Bassa$^{2}$ 
\\ $^1$ Jodrell Bank Centre for Astrophysics, School of Physics and Astronomy, The University of Manchester, Manchester M13 9PL, UK
\\ $^2$ ASTRON, the Netherlands Institute for Radio Astronomy, Postbus 2, 7990 AA, Dwingeloo, the Netherlands
}

\title[Spin-down rate changes of PSR B0919$+$06]{Understanding the spin-down rate changes of PSR B0919$+$06}

\begin{document}

\maketitle

\begin{abstract}
We study the spin-down properties of PSR B0919$+$06 based on almost 30 years of radio observations. 
We confirm that the time derivative of the rotational frequency $\dot \nu$ is modulated quasi-periodically and show that it exhibits a repeating double-peaked structure throughout the entire observation span.
We model the $\dot \nu$ variation of the pulsar assuming two spin-down rates with sudden switches between them in time. Our results show that the double-peak structure in $\dot \nu$ has a repetition time of about 630 days until MJD  52000 (April 2001) and 550 days since then. During this cycle, the pulsar spin varies from the lower spin-down rate to the upper spin-down rate twice with different amounts of time spent in each state, resulting in a further quasi-stable secondary modulation of the two-state switching. This particular spin-down state switching is broadly consistent with free precession of the pulsar, however, a strong evidence linked with this mechanism is not clearly established. We also confirm that the pulsar occasionally emits groups of pulses which appear early in pulse phase, so-called "flares", and these events significantly contribute to the pulse profile shape. We find the $\dot \nu$ modulation and the pulse shape variations are correlated throughout the observations. However, the flare-state is not entirely responsible for this correlation.
In addition to the flare-state, we detect flare-like events from the pulsar in single pulse observations. During these events, the shift in pulse phase is small compared to that of the main flare-state and clearly visible only in single pulse observations.
\end{abstract}


\section{Introduction}

Pulsars are one of the most stable rotators in the universe and the stability of the rotation of some pulsars is comparable to that of an atomic clock \citep[e.g.][]{pt96}. Pulsars spin-down gradually due to the loss of rotational kinetic energy into high-energy plasma and electromagnetic radiation. A simple spin-down model based on the rotational frequency $\nu$ and its first time derivative $\dot \nu$ is sufficient to explain their slow-down in general. However, most pulsars show irregularities in their spin properties, mainly due to glitch events \citep[i.e. a sudden increase in rotational frequency; e.g.][]{els+11,ymh+13} and low-frequency timing noise. 
These spin irregularities are generally seen in pulsar timing residuals; i.e. the difference between the measured time-of-arrival (TOA) of the pulse from the pulsar at the observatory and the spin-down model predicted TOA.
\citet{hlk10} studied timing irregularities for 366 pulsars using long term observations, time-scales over 10 years, and reported that some older pulsars show a quasi-periodic behavior in their timing residuals, while the timing residuals for young pulsars are dominated by the recovery from glitch events.
By analyzing slow-down rates for 17 pulsars with more than 20 years of observations, \citet{lhk+10} found that the timing behavior of these pulsars mainly results from two different spin-down rates. Several of these pulsars show quasi-periodically varying $\dot \nu$ with long time-scales ($>$ 1 yr). The $\dot \nu$ variation for six of these pulsars are correlated with changes in the pulse profile shapes, mainly switching between two shapes.  
This implies that the $\dot \nu$ variations are linked with some phenomena in the pulsar magnetosphere. The abrupt pulse profile variation between two shapes (sometimes three or many) in time is called mode changing \citep[e.g.][]{bac70a,ran86,ksj13}. The spin-down interpretation is strengthened by the fact that the two states of the intermittent pulsars (i.e. the radio-loud `on' state and the radio-quiet `off' state) are attributed with different $\dot \nu$ states \citep[see][]{klo+06,llm+12,crc+12,ysl+13}. By analyzing approximately 13 years of observation, \citet{ysl+13} reported that the intermittent pulsar PSR B1931$+$24 has two $\dot \nu$ modes corresponding to its radio `on' and `off' states, respectively, confirming the result of \citet{klo+06}. Therefore, it is well established that many pulsars show spin irregularities in their observation, although the exact mechanism of spin-down state and profile mode switching is not clearly understood.

\citet{Jon12} proposed that free precession is a possible explanation for pulsar spin-down rate changes. Assuming the neutron star is a biaxial body, the dipole radiation of the pulsar varies due to precession in a smooth periodic behavior, and similarly the spin-down torque. Consequently, the particle accelerating electric field is modulated by this periodic behavior and thus the kinetic energy of the plasma in the magnetosphere.
However, the sharp switches between the spin-down rates are likely associated with a capacitor-like process in the magnetosphere, overlapping with the smooth periodic behavior mentioned above. 
He explains that the pair production is a clear candidate for this capacitor-like process. 
In this process, the high-energy photons radiated by an accelerating plasma can produce electron-positron pairs when the photon energy exceeds twice the rest mass energy of an electron. These pairs may produce further photons and then further pairs. In general, it is believed that these pairs are responsible for the radio emission of pulsars \citep[e.g.][]{rs75,dh86,hm02,jon14}. If the plasma kinetic energy exceeds the required minimum energy for this process, the pair production begins and the magnetosphere switches from the low spin-down state to the high spin-down state abruptly, and vice-versa. 
The model given in \citet{alw06} explains a triaxial body under free precession is capable of generating a more complicated quasi-periodic timing residuals like those seen from PSR B1828--11 \citep[see][for more detail]{lhk+10}.

Based on two different magnetospheric configurations with plasma availability in the magnetosphere, \citet{lst12a} proposed a mechanism for $\dot \nu$ variation and mode changing of intermittent pulsars.
They assumed the plasma-filled force-free limit for the `on' state and the vacuum limit for the `off' state within the open field line regions to explain the corresponding two spin-down rates. They found the ratio of the two spin-down rates between `on' and `off' states to be $\sim$1.2--2.9, which is consistent with observation. However, the existence of these two ideal limits in a real pulsar magnetosphere is doubtful. If the magnetosphere is in the force-free limit, then the parallel electric field component to the magnetic field line is screened. Therefore, the particle acceleration cannot take place in the magnetosphere, so that the electromagnetic radiation cannot exist. 
On the other hand, maintaining a vacuum condition in the magnetosphere is highly unlikely due to the presence of a non-zero electric field component perpendicular to the neutron star surface. If the vacuum condition is maintained in the magnetosphere, then there are no free particles to accelerate and produce electromagnetic radiation. 
Therefore, as described in recent studies, the radio emission mechanism in pulsar magnetosphere is likely to be operated in between these two ideal limits \citep[see][]{lst+12,kkh+12,khk12a}.

\citet{sl13} proposed a technique to resample a pulsar time series evenly, thereby retaining the same information measured in the unevenly sampled series. By applying this technique to the pulsars given in \citet{lhk+10}, they demonstrated that the pulsar spin down rate variations may be chaotic in nature. They emphasized that PSR B1828--11 exhibits a clear chaotic behavior and explained that it is due to spin down rate variation of the pulsar and not caused by random processes.

In this work, we study the $\dot \nu$ variation from one of the $\dot \nu$ state-changing pulsars reported in \citet{lhk+10}, PSR B0919$+$06, in detail. Here we report the first study of pulse-shape related $\dot \nu$ variation for this pulsar. PSR B0919$+$06 is a relatively bright radio pulsar with a rotational frequency of $\nu = 2.3$~Hz and a slowdown rate of $\dot \nu = -7.4\times10^{-14}$~Hz s$^{-1}$ \citep{mlt+78}. Due to the high flux density, the single pulses from the pulsar are clearly visible. 
By closely studying the single pulses of the pulsar, \citet{rrw06} found that occasionally the emission appears early in pulse longitude compared to the normal emission, about $0.015$ in pulse phase, and this mode lasts for about 5--15~s; hereafter we name such an event as a `flare-state'.
The transition from the normal emission to the flare-state occurs gradually, and remains in that state for typically several tens of pulses, and then gradually returns to the normal emission mode.
According to their study, the flare-state is rare in time, typically one in several thousand pulses.
Recently, \citet{Sha10} studied the modulation in timing residuals and $\dot \nu$ variation in time and explained that the pulsar had undergone a continuous sequence of 12 spin-down rate variation cycles within the time between 1991 and 2009, confirming the result of \citet{lhk+10}.
She described the observed frequency residuals $\Delta \nu$ relative to the timing model due to $\dot \nu$ variation with a sawtooth-like function with a periodicity of 600 days, similar to what has been found by \citet{lhk+10}.
Further, she reported a large glitch event occurred from this pulsar on 5 November 2009 (MJD 55140), which caused a fractional increase in spin frequency of $\Delta \nu / \nu \sim 1.3 \times 10^{-6}$.
Since we have started recording the high-quality data with the new pulsar backend right after the glitch, we analyze the data span separately before and after the glitch epoch MJD $55140$ in this work.

In Section~\ref{obs}, we present our observations, data processing and analysis. We analyze pulse profiles using sub-integrations in time and identify the flare-state in Section~\ref{prof}. The spin-down states and pulse profile shape variations are presented in Section~\ref{spin_down}. In addition to the flare-state, we identify flare-like events in single pulse observations and present these in Section~\ref{single}. In Section~\ref{sim}, we simulate the observed spin-down modulation using two $\dot \nu$ rates. Finally, we discuss and summarize our results in Section~\ref{dis}.

\section{Observations}
\label{obs}

We observed PSR B0919$+$06 using the 76-m Lovell Telescope at the Jodrell Bank Observatory since 1984 August 30. We also used the 25-m MKII telescope located on the Jodrell Bank site to observe the pulsar occasionally. 
The data were collected from three different pulsar backends: `Analog-filterbank (AFB)', `Digital-filterbank (DFB)', and `ROACH', depending on when the observation was made. Note that the resolution of the AFB data is low compared to that of the DFB and ROACH data. In our analysis, we used L-band/1400-MHz data collected from the AFB until the glitch event occurred in November 2009 and the DFB since then. The observations are about 6 min long with different sub-integration lengths: AFB data are recorded with 1-min sub-integrations, and DFB data with 10-s sub-integrations. Each sub-integration represents a pulse profile formed by integrating all single pulses recorded within the sub-integration length. Since the length of the flare-state is $\sim$5--15 s and similar to the sub-integration length of the DFB data, we used DFB data in the analysis of the pulse profiles of the flare-state and non-flare state. Due to this sub-integration time-length, we note that sometimes the flare-state appeared in three adjacent sub-integrations. There were few observations made between 1984 August 30 and 1989 April 9, and they were of poor quality. Thus, we ignore these data in our analysis.

In addition to the DFB data with 10-s sub-integrations, we simultaneously used the backend ROACH since June 2011 and recorded single pulse data. We use the single pulse data to investigate the characteristics of the flare- and flare-like states in detail in Section~\ref{single}. We carried out 20-min long observations once every week since 24 November 2013 and used both the DFB (with 10-s sub-integrations) and ROACH (with single pulses) to record the data.

\section{Pulse profiles of normal- and flare-states}
\label{prof}

\begin{figure}
\begin{center}
\psfig{file=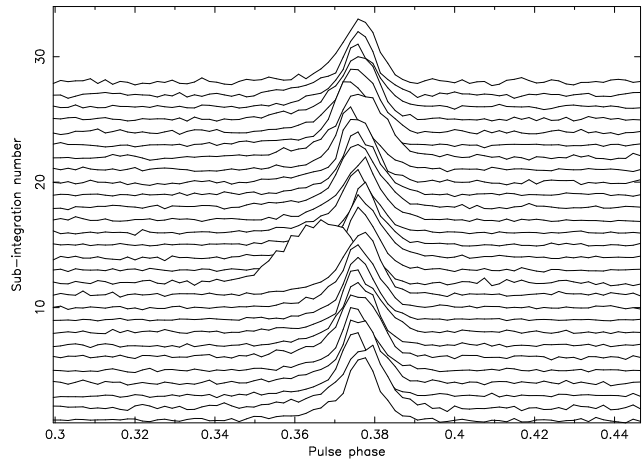,width=3.3in,angle=-90}
\end{center}
\caption{
The profiles of the 10-s sub-integrations on MJD 55165. 
The flare-state occurred within sub-integration number 12, where the radio pulse appeared early in pulse phase. Only a fraction of the full range of pulse phase is shown. 
\label{stack}}
\end{figure}

\begin{figure}
\psfig{file=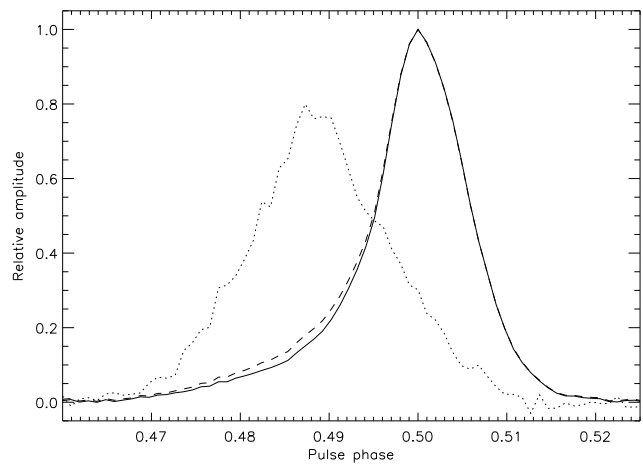,width=3.3in}
\caption{
The average pulse profiles of the normal- ({\it Solid}) and flare-state ({\it Dotted}) of PSR B0919$+$06. There are 1024 bins across each profile in pulse phase and only a fraction of the full range of pulse phase is shown. Note that the peak of the normal-state profile is normalized to unity and the flare-state profile is scaled according to the peak flux density ratio (see the text).
The flare-included average pulse profile is shown as a {\it Dashed} line.
\label{profiles}}
\end{figure}

We use the DFB data to construct the pulse profile in the two states. By comparing the profiles of sub-integrations in each observation, we easily identify the flare-state since the pulse appears early in pulse phase during this state compared to the most common normal-state. Figure~\ref{stack} shows the sub-integrations on MJD 55165 with the flare-state appearing within sub-integration number 12. We separate the sub-integrations containing flares and then sum them together to form the average pulse profile of the flare-state, while we use the rest of the sub-integrations to form the average pulse profiles of the most common normal-state (Figure~\ref{profiles}). In the normal-state pulse profile, approximately 1090 sub-integrations were used, resulting in about a total of approximately 3 hours of data. In contrast, we used 37 sub-integrations of the flare-state corresponding to about 6~min of data. Due to this large difference in data lengths, the normal-state profile has about 6 times higher signal-to-noise ratio. Further, the ratio between the peak flux densities of the normal-state and the flare-state profiles is calculated to be about 1.2, and the ratio of the area under the pulse (i.e. the power) of the two profiles is almost unity. 
With our 10-s sub-integration length, the pulse profile of the flare-state may be slightly mixed with the normal-state profiles since the flare-state may have appeared partially in 2--3 adjacent sub-integrations.

Within 13 recent 20-min long observations, we identified at least one flare-state in every observation and two flare-states in three observations. The repetition time-scale of flare-states from these three observations is calculated to vary between about 1000--1850 pulses. However, this is a crude estimate and we do not have enough data to make a precise periodicity calculation.

\section{Spin-down states and pulse profile shape parameters}
\label{spin_down}

\begin{figure*}
\centering\psfig{file=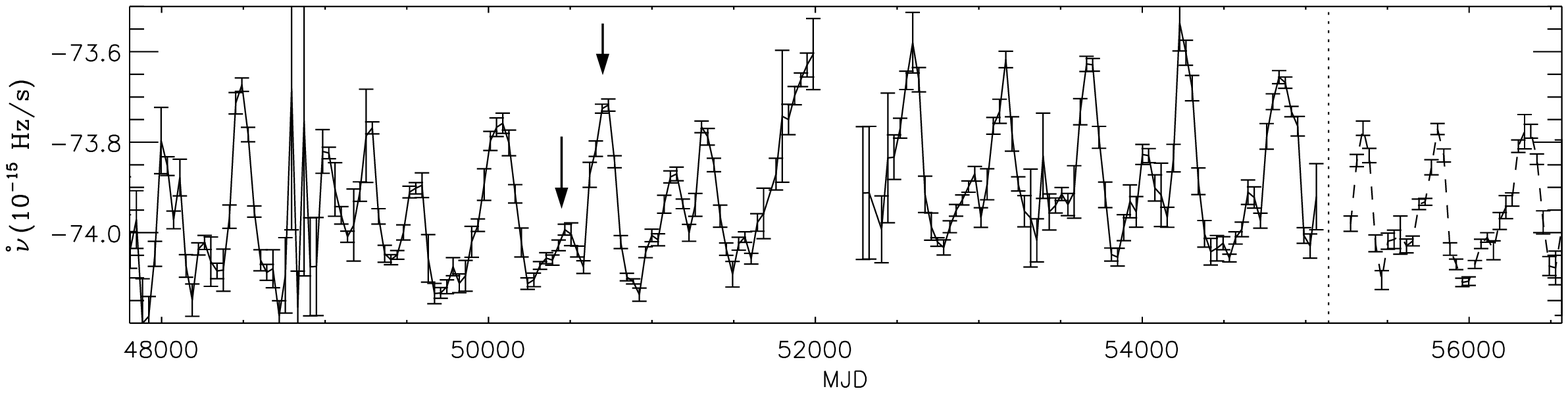,width=7in}
\caption{
The $\dot \nu$ variation over time for the data obtained before the glitch epoch ({\it solid}) and after the glitch epoch ({\it dashed}).
A double-peak like structure (a small peak followed by large peak -- arrows indicate an example of the two peaks) repeats over time throughout the entire observation span.
Note that the gap around MJD 52000 is due to an extended maintenance period of the Lovell Telescope. The {\it dotted} line represents the glitch epoch MJD 55140.
\label{nu_all}}
\end{figure*}

In order to study the different spin-down states and the variation in $\dot \nu$ over time, we find timing solutions for subsequent partially overlapping sections of data. To be consistent with \citet{lhk+10}, we first use $T=150$~day sections with $T/4$ ($=38$~day) strides along the entire data set. There is an average of about 20 observations per 150~day section.
For each section, we fit for $\nu$ and $\dot \nu$ using the pulsar timing package {\sc psrtime}\footnote{http://www.jb.man.ac.uk/pulsar/observing/progs/psrtime.html}, and use this method for the data before and after the glitch epoch MJD 55140 separately. In addition to $T = 150$~day sections, we use $T=100$~day and 80-day sections with $T/4$ strides to study the modulation in $\dot \nu$ with short-length data sections. 
We find that all these fits result in a similar modulation pattern of $\dot \nu$. Since the $T = 150$~day fit provides a smooth variation in the modulation, we use this particular choice throughout this work.
Our results are shown in Figure~\ref{nu_all} for both pre- and post-glitch data. 
Note that the gap right after about MJD 52000 is due to lack of observations around these days when a telescope renovation was taking place; the pulsar was not observed in 2001 between March and December.


As shown in Figure~\ref{nu_all}, a double-peak structure (a large peak followed by a small peak) is clearly seen and it repeats over time, almost throughout the entire observation span of 30 years. \citet{lhk+10} explained that the $\dot \nu$ variation of pulsars over time is associated with two different spin-down states and the switching time-scale between states is very short. Adopting this model for PSR B0919$+$06, we investigate and model the $\dot \nu$ variation with two spin-down states in Section~\ref{sim}.

\begin{figure}
\psfig{file=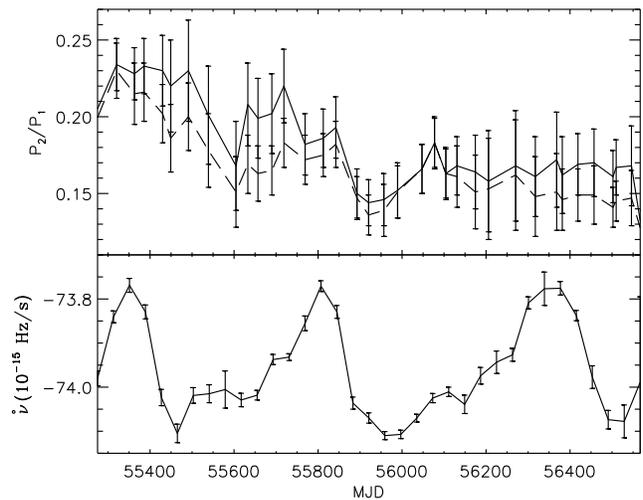,width=3.3in}
\caption{
The evolution of the averaged shape parameter including ({\it solid}) and excluding ({\it dashed}) flares after the epoch MJD 55140. We plot the variation in $\dot \nu$ over time in the bottom panel for comparison.
\label{shape_flare}}
\end{figure}

\begin{figure}
\psfig{file=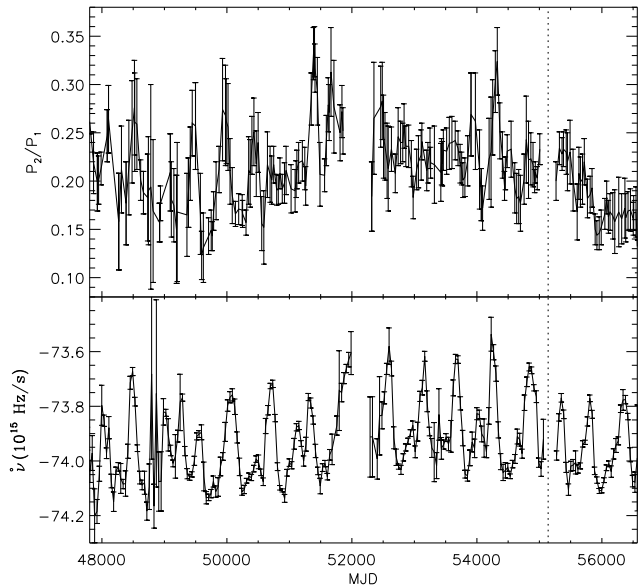,width=3.3in}
\caption{
The evolution of averaged shape parameter for the entire data span, including the low-resolution data obtained until the glitch epoch and the high-resolution data since then. The gap around MJD 52000 is due to an extended maintenance of the telescope. The {\it dotted} line shows the glitch epoch.
\label{shape_flare_pre}}
\end{figure}

In order to see any correlation between pulse profile shapes and the $\dot \nu$ variation, as shown in \citet{lhk+10} for six pulsars, we study the variation in pulse profile shape parameters over time. 
We first obtain pulse profiles for each observation and then synthesize these profiles by fitting two gaussians to eliminate high frequency noise. Then we use this synthetic profile to measure all the shape parameters, including pulse width at different intensity levels, amplitude and power of the two gaussians, etc. 
We find that all the shape parameters follow a similar modulation result. Since the ratio of powers (i.e. the areas) of the two gaussians represents the profile variation better compared to the other shape measurements, we use this throughout our analysis as the shape parameter of the pulse profile. 
We then average our power ratio measurements according to the previously used data lengths (i.e. $T=150$~d with 38~d strides).
The uncertainties of these power ratio measurements are calculated from the standard deviation of the ratios obtained from the individual pulse profiles.
The AFB data has a resolution of 512 bins across the pulse phase of the pulse profile. Therefore, we rebin the pulse profiles of the DFB data to 512 bins to be consistent with the AFB data.
In order to investigate the involvement of the flare-state on the $\dot \nu$ evolution, we analyze the shape parameter obtained from pulse profiles including and excluding flares separately.   
Figure~\ref{shape_flare} shows the variation in shape parameter, the ratio of the power of the two gaussians, over time after the glitch epoch. 
It is clear that the shape parameter obtained from flare-subtracted pulse profiles is lower than the shape parameter of the flare-included profiles, as expected from Figure~\ref{profiles}. 
We plot the $\dot \nu$ variation in the same figure for comparison.
We then follow the same analysis for pre-glitch data including flares. Note, that we do not measure flare-subtracted shape parameters for pre-glitch data, because these data used 1~min long sub-integrations, so that it is not possible to separate flare-state emission. The shape parameter result is shown in Figure~\ref{shape_flare_pre}.

\bigskip

\begin{figure}
\psfig{file=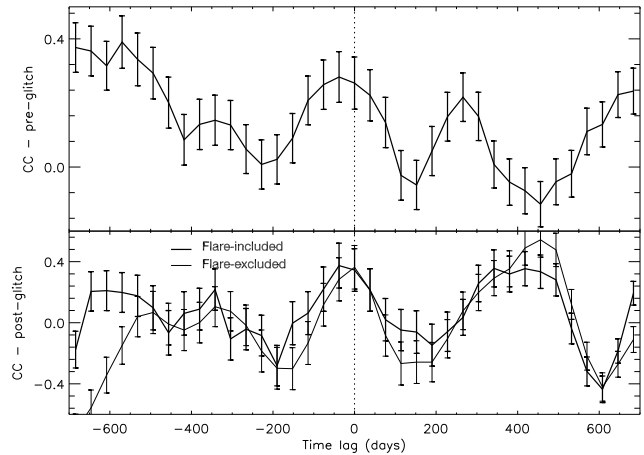,width=3.3in}
\caption{
The cross-correlation coefficients between $\dot \nu$ and the shape parameter (i.e. the ratio of the power of the two gaussians). The {\it top} panel shows the coefficients for pre-glitch data (i.e. the low-quality data). The {\it bottom} panel shows the coefficients for post-glitch data (i.e. high-quality data) including ({\it thick solid}) and excluding ({\it thin solid}) flares. 
\label{crosscor}}
\end{figure}

In order to examine any correlation, we cross-correlate the measured $\dot \nu$ with the shape parameter.
Some early observations were noisy and we removed them in the shape parameter calculation.
To calculate the uncertainties of the coefficients, we randomize the order of the measured shape parameters and cross-correlate with the measured $\dot \nu$. We perform this analysis for 1000 trials and then quote the standard deviation of the coefficients as the uncertainty of each time lag.
We find that the cross-correlation coefficient (at zero time lag) for flare-included and flare-subtracted shape parameters obtained from post-glitch high-quality data are $0.35\pm0.14$ and $0.36\pm0.14$, respectively, resulting in about a $2.5\sigma$ correlation. The correlation coefficient for pre-glitch low-quality (but much longer time span) data is calculated to be $0.26\pm0.08$, resulting in a $3.3\sigma$ correlation. To examine the correlation across the entire data span, we combine the two data sets together regardless of the data quality and then calculate the cross-correlation. We find the correlation coefficient for this case is to be $0.33\pm0.07$, confirming a $4.7\sigma$ correlation.
We note that the derived errors are somewhat overestimated, making the quoted significances conservative.
In Figure~\ref{crosscor}, we show the cross-correlation coefficient as a function of time lag between $\dot \nu$ and shape parameter curves. This figure shows that the shape parameter and $\dot \nu$ are periodic.
The noise of the shape parameter due to under-sampling may have caused the side-lobs and the slight offset of the correlation peak from the zero time lag. Further this shows that the shape parameter measured from both pre- and post-glitch data sets are correlated with $\dot \nu$. If the flare-state is fully associated with the $\dot \nu$ changes, then we should not see any correlation between $\dot \nu$ and shape parameter obtained from flare-subtracted pulse profiles. However, the correlation seen in our results (see Figure~\ref{crosscor} bottom panel) indicates that there is no strong link between the flares and $\dot \nu$ changes. In order to investigate any hidden flares in our 10-s sub-integrations, we use the single pulse data as described in Section~\ref{single}.

\begin{figure}
\psfig{file=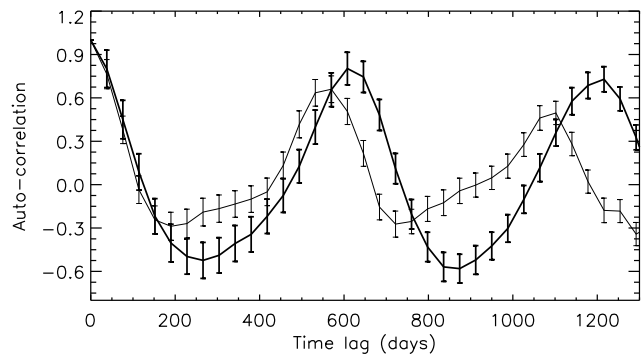,width=3.3in}
\caption{
The auto-correlation function for the data obtained before MJD 52000 ({\it thick solid}) and after MJD 52000 ({\it thin solid}). Note that the two curves show the periodicity of the $\dot \nu$ modulation varied from about 630 days to 550 days around MJD 52000.
\label{auto}}
\end{figure}

We then investigate the periodicity of the double-peak structure of the $\dot \nu$ variation curve. We notice that there is a shift in the modulation pattern of the $\dot \nu$ around MJD 52000. Therefore, we calculate the auto-correlation function for the data obtained before and after MJD 52000 separately, and present in Figure~\ref{auto}. This shows that the periodicity for the data obtained before and after MJD 52000 are about 630 days and 550 days, respectively (consistent with the periodicity seen in Figure~\ref{crosscor}). We further study this shift in Section~\ref{sim} with the simulation of the $\dot \nu$ variation.

\begin{figure}
\psfig{file=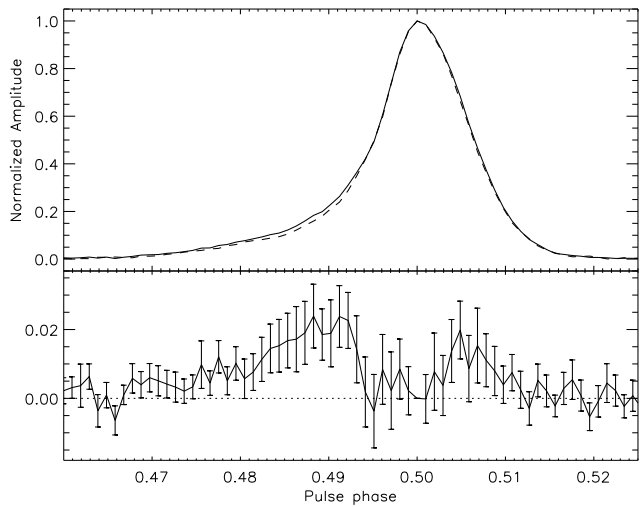,width=3.3in}
\caption{
The two pulse profile modes corresponding to low ({\it solid}) and high ({\it dashed}) $\dot \nu$ states for the data obtained after the epoch MJD 55140. In order to see the profile shape difference, we plot a fraction of the pulse phase. The bottom panel shows the difference between the two profiles which are normalized such that the peak has an amplitude of unity.
\label{prof_modes}}
\end{figure}

Regardless of the flare-state, Figure~\ref{crosscor} shows that the pulse profile shapes are correlated with spin-down rates. Therefore, we expect to have two different pulse profile shapes corresponding to the high and low spin-down rates. 
In order to examine this with high-resolution post-glitch data, we adopt the high spin-down state when $\dot \nu > -7.394 \times 10^{-14}$~Hz s$^{-1}$ and the low spin-down state when $\dot \nu < -7.394 \times 10^{-14}$~Hz s$^{-1}$. 
We determine this cut-off value by averaging the overall $\dot \nu$ across the post-glitch data span given in Figure~\ref{shape_flare}. Note that, a dynamically varying cut-off is suitable for pre-glitch data due to the varying mean of $\dot \nu$ over time (see Figure~\ref{shape_flare_pre}), but we do not attempt to use this data section for this particular analysis due to its lower quality. Then we find the two integrated pulse profiles corresponding to the two spin-down states for post-glitch data. Figure~\ref{prof_modes} shows the two pulse profiles. 
The uncertainties of the difference of the two profiles are calculated by randomizing the two sets of individual pulse profiles and then calculating the standard deviation of the intensity difference in each pulse phase bin based on 1000 trials.
It is seen that the pulsar has a slightly narrower pulse profile when it is in the high spin-down state compared to when it is in the low spin-down state. This confirms that there are two pulse profile modes, however, the difference between them is not as significant as the mode change in some other pulsars \citep[see][]{ran86,wmj07,lhk+10}.

\begin{figure}
\psfig{file=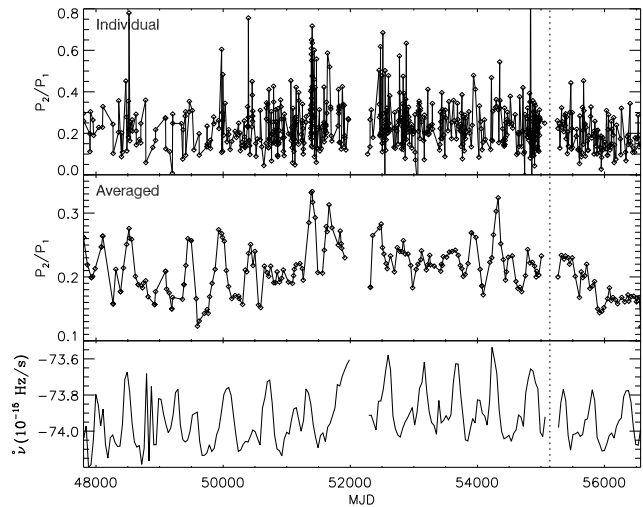,width=3.3in}
\caption{
The evolution of the shape parameter (the ratio of the power of the two gaussians) for individual ({\it top panel}) and averaged ({\it middle panel}) observations. Note that the bottom panel shows the $\dot \nu$ variation. The {\it dotted} line shows the glitch epoch.
\label{indi}}
\end{figure}

As given in \citet{lhk+10} for PSR B1828--11, the $\dot \nu$ variation can be correlated with shape parameters obtained from individual observations. Therefore, in order to study such a correlation on short time-scales, we analyze the shape parameter from PSR B0919$+$06 for individual observations. For comparison, we plot the $\dot \nu$ variation with these individual observation shape parameter of power ratio of the two gaussians in Figure~\ref{indi}. This shows that the modulation pattern of the ratio from individual observations is complicated. It is likely that the pulsar switches between modes rapidly on very short time-scales ($<$1~day). Furthermore, we cannot identify the time-scale of this behavior clearly due to under-sampling of the time-scale. This suggests that it is likely the pulsar spin-down switches between two modes rapidly while it spends more time in one mode compared to the other mode for a given period of time. 
This is why we obtained a long term modulation pattern in both averaged $\dot \nu$ and shape parameter curves (see Figure~\ref{shape_flare} and \ref{shape_flare_pre}).

\section{Flare-state in single pulse observations}
\label{single}

\begin{figure}
\psfig{file=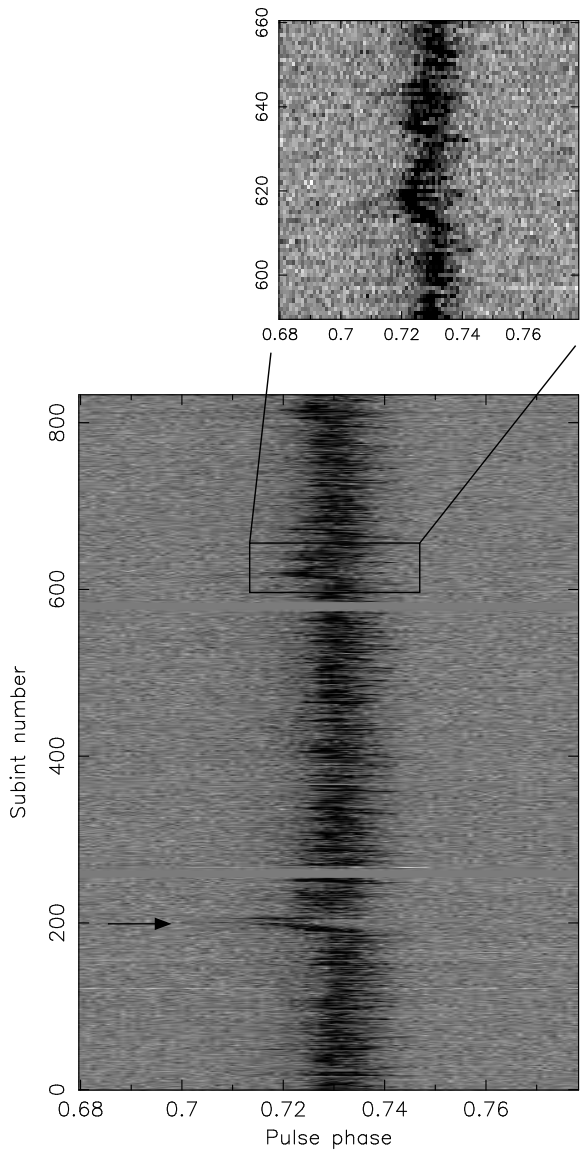,width=3in}
\caption{
The single pulse data on MJD 55751 (2011 July 9). A flare-state can be seen around the single pulse index 200, which is indicated with an arrow. A flare-like event occurs around index 620 and a zoomed window of this region is shown in the inset. 
\label{flare2d}}
\end{figure}

As shown in the previous section, the correlation between the $\dot \nu$ variation and flare-state is not clearly evident with 10-s sub-integrations. Therefore, we investigate the single pulse data obtained from the ROACH since June 2011. We clearly identify the flare-state in these data. In addition to this main flare-state, we find some events in which the single pulses appear slightly early in pulse phase with respect to the normal emission, but the shift is about a factor of two smaller compared to that of the main flare-state.
Also, the duration of these events is about a factor of two shorter than that of the main flare-state.
Figure~\ref{flare2d} shows the single pulse data on MJD 55751. The flare-state occurs around pulse 200 and also note that there is a flare-like event around pulse number 620. 
As can be seen from the inset, this event is likely a small flare and is not clearly visible in the 10-s sub-integration data. 
Therefore, these have not been removed in the flare-subtracted data set which we used in the previous section to make correlation results. 
These events can be a weak phase of the main flare-state, or an additional small flare-state. We see these type of events in a few other single pulse observations obtained on different days.
The difference between the pulse profiles of the two modes corresponding to high and low $\dot \nu$ states shown in Figure~\ref{prof_modes} could be due to these flare-like events. In order to investigate the importance of these events for mode changing, we need a detailed analysis using long-term single pulse observations.

\section{Simulating the spin-down modulation}
\label{sim}

\begin{figure}
\psfig{file=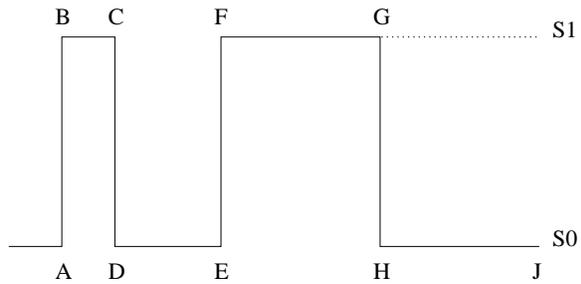,width=3.in}
\caption{
Schematic diagram of the intrinsic mode switching of the pulsar between two spin-down states across the 550-day cycle. The horizontal axis represents days and the switching between the two states occurs suddenly in a very short time-scale. The pulsar switches from the lower spin-down rate state (S0) to the upper spin-down rate state (S1) twice during the cycle.  
\label{diag}}
\end{figure}

In this section, we attempt to simulate the measured modulation in $\dot \nu$ over time with several spin-down rate states. 
Although Figure~\ref{nu_all} might suggest the existence of three $\dot \nu$ rate states, we show here that it is possible to model the observed $\dot \nu$ pattern using only two spin-down rate states.
We denote the high spin-down state as `S1' and the low spin-down state as `S0'. 
As shown in Figure~\ref{nu_all}, we identified a double-peak structure in the $\dot \nu$ modulation, a smaller peak coupled with a larger peak, and this repeats over time.
To simplify the model, we ignore the rapid switching observed in the shape parameter from the individual observations (Figure~\ref{indi}) and define the average behavior as the state switching of the pulsar. The auto-correlation results given in Section~\ref{spin_down} showed that the period of this repeating structure is about 550 days for the post-glitch data where the data quality is higher. 
This period is consistent with the period of the observed sawtooth-like curve of $\Delta \nu$ given in \citet{Sha10}.
Figure~\ref{diag} shows a schematic diagram of the model we adopt for modeling the spin-down state switching across the double-peak structure. Thus, the two different heights of the observed $\dot \nu$ peaks shown in Figure~\ref{nu_all} is interpreted as different mode durations with the two intrinsic $\dot \nu$ states across the entire cycle of 550 days (see Figure~\ref{diag}).

In order to simulate the observed modulation in $\dot \nu$ over time, we first generate fake TOAs for a model-pulsar which has the same pulse period as that of B0919$+$06. 
We use two $\dot \nu$ values of $-7.4\times 10^{-14}$~Hz/s and $-7.36\times 10^{-14}$~Hz/s for S0 and S1, respectively. 
Then we generate TOAs assuming a typical uncertainty of 150~$\mu$s with equal sampling of 5 days and assign the two spin-down states as shown in Figure~\ref{diag} for the given time-scales in each state.
We use the pulsar timing software developed in \citet{wje11} for the glitch study of PSR J1119--6127.
Finally, we find $\dot \nu$ values from these simulated TOAs using average-stride fits using the same procedure performed on real data. By matching the simulated and observed $\dot \nu$ curves, we determine the time-scales the pulsar spends in each state across the double-peak structure.

\begin{figure*}
\psfig{file=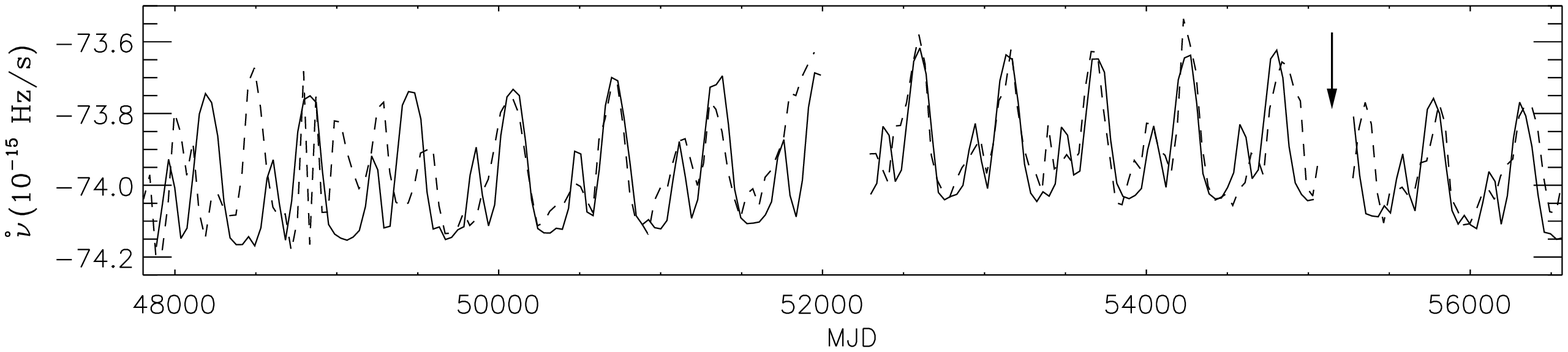,width=7in}
\caption{
The simulated ({\it solid}) and observed ({\it dashed}) $\dot \nu$ variations over time. The cycle of the double peak structure for the simulated data before and after MJD 52000 is 630 days and 550 days, respectively. The glitch event occurred on MJD 55140 and this epoch is marked by an arrow. Note that due to poor quality of early data (i.e., before MJD 49500), the double-peak structure is not clearly evident.  
\label{f1_sim}}
\end{figure*}

We study low-resolution pre-glitch and high-resolution post-glitch data separately. As mentioned in Section~\ref{spin_down}, we find that there is a shift in the modulation pattern of $\dot \nu$ around MJD 52000. Therefore, we match the simulated curve with data separately for before and after MJD 52000 within the pre-glitch data span. 
The period of the cycle of the double-peak structure in $\dot \nu$ before MJD 52000 is about $630$ days.
Figure~\ref{f1_sim} shows the simulated $\dot \nu$ curve for the data, over-plotted with the observed curve. 
The time-scales in the simulated mode switching between the two $\dot \nu$ states across the double-peak structure are given in Table~1 (see Figure~\ref{diag} for notation). 
We note that the low quality data obtained in early observations results in a poor quality $\dot \nu$ curve before about MJD 49500. Thus, we mainly focus on the $\dot \nu$ variation after MJD 49500 when we match the simulation with observation.

\begin{table} 
\label{param}
\begin{center}
\caption{
The time-scales in the simulated mode switching between the two $\dot \nu$ states across the double-peak structure for before and after MJD 52000. The values are in days and the notation are given in Figure~\ref{diag}.
}
\begin{tabular}{lllll}
\hline
\multicolumn{1}{c}{} &
\multicolumn{1}{c}{AJ} & 
\multicolumn{1}{c}{BC} &
\multicolumn{1}{c}{DE} & 
\multicolumn{1}{c}{FG} \\
\hline 
Before MJD 52000 & 630 & 50 & 120 & 175 \\ 
After MJD 52000 & 550 & 44 & 105 & 153 \\
\hline \end{tabular}
\end{center}\end{table}

We then followed the same analysis for the data section between  MJD 52000 and the glitch event, and found that the above time-scales for mode switching do not provide a good fit. 
The double-peak structure within this section of data reveals that the period of its cycle is about 550 days, similar to the post-glitch data (see Table~1). 
Therefore, we rescale the time-scales of the simulated $\dot \nu$ state switching that we used for the data section before MJD 52000 by a factor of 0.87 (see Table~1). Then we find that the simulated $\dot \nu$ modulation fits the observed curve better (see Figure~\ref{f1_sim}).  
However, we apply a phase shift of 80 days between the two simulated curves corresponding to before and after the epoch MJD $52000$ in order to match with the observations.

Finally, we simulate the post-glitch data following the same method given above for the pre-glitch data. We find that the simulated $\dot \nu$ variation is capable of explaining the observed curve with the same time-scales that we used in the data between MJD 52000 and the glitch (see Table~1). The simulated and observed curves are shown in Figure~\ref{f1_sim}. This suggests that the time scales of state switching were somehow retained during the glitch event. 
However, the phase of the cycle is off by 100 days after the glitch and we corrected this in Figure ~\ref{f1_sim}.
We also note that there is a slight offset between the simulation and the observation right before and after the glitch. Further, we note that there is a slight variation in averaged $\dot \nu$ measurements over time with a varying overall slope and can clearly be seen in Figure~\ref{nu_all}. In order to match this variation in the slope of $\dot \nu$, we assumed a positive, zero, and negative $\ddot \nu$ values in the simulation for the data sections before MJD 52000, between MJD 52000 and the glitch, and post-glitch, respectively. However, we note that this variation in slope is not necessarily due to a stable $\ddot \nu$ (hence a reliable measurement of the braking index) and the exact reason is not yet clearly understood.

With the lack of observations made around MJD 52000 due to telescope renovation, we were unable to study the phase shift in the $\dot \nu$ modulation that happened around this time. However, by comparing with the observation reported in \citet{Sha10}, it is clear that the pulsar has not undergone a glitch around this epoch. Therefore, the phase shift of 80 days in our simulation before and after this epoch reflects that there was a modification in the pulsar spin-down modulation itself somehow, independent of a glitch event. 
This phenomenon also changed the time-scale of the double-peak structure in $\dot \nu$ from 630 days to 550 days. However, the cause of this modification in pulsar spin-down modulation is not easily explainable.
This shift is also visible in the data reported in \citet{Sha10} (see Figure~4 therein) collected from the Pushchino Observatory. Therefore, it is clear that this modification in the periodicity of the modulation pattern is not introduced by the way that the data were analyzed.

\section{Summary}
\label{dis}

Pulsars generally show spin irregularities in their timing behavior.
As reported in \citet{lhk+10}, the correlation between $\dot \nu$ variations and pulse profile shapes of some pulsars suggests the $\dot \nu$ variation is a consequence of a magnetospheric effect.  
The shape parameters of PSR B1828--11 for individual observations show that the mode switching can occur rapidly within a very short time-scale \citep[see Figure~5 in][]{lhk+10}. 
However, the average shape parameters obtained from stride fitting analysis show a slowly varying periodic pattern with a relatively large time-scale.
The mechanism responsible for these correlated changes is not clearly understood.
\citet{alw06} and \citet{Jon12} proposed free precession of the pulsar is a possible mechanism for $\dot \nu$ related periodic mode changing.

In this work, we analyzed the pulse profile shape variation and the spin-down rate changes of PSR B$0919+06$ in detail using almost 30 years of observation from the Lovell and MKII Telescopes in the Jodrell Bank Observatory.
As first reported in \citet{rrw06}, we identified the flare emission from the pulsar that occurs occasionally along with its usual radio emission. 
Using pulsar timing, we found that the spin-down $\dot \nu$ of the pulsar varies quasi-periodically, as previously noted in \citet{lhk+10} and \citet{Sha10}, exhibiting a double-peak modulation structure in time (see Figure~\ref{nu_all}) over the entire observation. This is more similar to the $\dot \nu$ variation of PSR B1828--11 \citep[see Figure~2 in][]{lhk+10}.
We then measured the average pulse profile shape parameter (i.e. the ratio of the power of the two gaussians fitted to the observed pulsar profile) for sections of data along the 30 years and found that the pulse profile shape varies over time. In order to investigate the influence from flares on the modulation in pulse profile shape variation, we subtracted the identified flare-states in our data and then re-measured the profile shape parameter. 
We found that the flare-state has a direct impact on the shape parameter.
By cross-correlating the variations in $\dot \nu$ and the shape parameter, we found that the modulation of these two parameters are correlated. This correlation exists even when the flare-state has been removed from the data. 
This suggests that the flare-state is not necessarily related to the observed $\dot \nu$ variation of the pulsar and if it is linked, it can only partially account for the observed correlation.

Using single pulse observations, we identified flare-like events in addition to the main flare-state. These events are shorter in duration, about factor of two smaller than that of the previously identified flare state (i.e., about 5~s), and coupled with smaller shifts in pulse phase, about $0.008$ in pulse phase. These events are more likely the weak phases of the flare-state. Therefore, they are not clearly visible in the data already averaged into 10-s sub-integrations. 
They may be associated with the profile variations seen in the pulsar that apparently correlate with $\dot \nu$ variations. They may also represent a continuum of flare types. It is not possible to check the relationship between flares and the $\dot \nu$ variations as they require single pulse observations to resolve them and we have been recording single pulses relatively recently.
In the future, we will be able to analyze only a longer single pulse data span and then determine the correlation between the $\dot \nu$ related mode changing and all-flare-states of the pulsar.

The measured $\dot \nu$ variation pattern of the pulsar shown in Figure~\ref{nu_all} might suggest the existence of three or more $\dot \nu$ states. However, we show that it is possible to model the $\dot \nu$ variation of the pulsar using only two distinct $\dot \nu$ values.
According to our results, the identified repeating double-peak structure in the $\dot \nu$ curve has a periodicity of 630 days and 550 days before and after the epoch around MJD 52000, respectively. 
Within this cycle, the pulsar spin-down switches from the low spin-down rate to the high spin-down rate twice (see Figure~\ref{diag}), resulting in a double-peak structure in the average $\dot \nu$ variation (see Figure~\ref{f1_sim}). The relative heights of the two peaks are different due to different amounts of time the pulsar spends in the two $\dot \nu$ states during the cycle (i.e., a shorter time in the first high $\dot \nu$ rate compared to the second transition; see Table~1). 
The change in duration of this longer modulation period from 630 days to 550 days happened at a time of no observations, but comparison with \citet{Sha10} seems to show that the change may have happened gradually. We also note that there seems to be a phase shift in the modulation at the time of the glitch which might hint at some relationship.
The mechanism which regulates this longer modulation cycle is not understood. Although this model avoids the necessity of more than two distinct spin-down values, it does require a mechanism explaining the alternate switches between a shorter and a longer high $\dot \nu$ states, resulting in a further secondary modulation of the two-state switching.

Free precession of the pulsar is possibly an explanation for this $\dot \nu$ variation. Then the repeating structure in $\dot \nu$ is due to precession and the time-scale of this cycle is the precession cycle. Although, the spin-down switching between two $\dot \nu$ rates with different time-scales in each state within the precession cycle cannot be easily understood. A possible explanation is that the locations and the shape of the pulsar radio emission region may vary slightly with the precession due to wobbling of the pulsar with respect to the total angular momentum axis. Thus, the pair production somehow enhances during two phases across the precession period with different durations and causes the double-peak structure in the $\dot \nu$ state switching.
This should provide a profile variation during the precession cycle and this is consistent with our shape parameter analysis.

Regardless of the exact $\dot \nu$ changing mechanism, we can generally explain any periodic averaged-spin-down modulation pattern using only two spin-down rates.
However, in order to understand the exact $\dot \nu$ and mode changing of the pulsar on short time-scales, we need an extensive analysis based on high resolution observation with almost equally sampled long-term single pulse data.

\noindent

\bigskip

\section*{Acknowledgments}
Pulsar research at the Jodrell Bank Centre for Astrophysics and the observations using the Lovell Telescope is supported by a consolidated grant from the STFC in the UK.

\bibliography{psrrefs,modrefs,journals,0737Ack}
\bibliographystyle{mn2e}

\end{document}